\documentclass[twocolumn,prb,showpacs,multicol,amsmath,amssymb]{revtex4}
\usepackage[dvips]{graphicx}
\usepackage{graphicx}
\usepackage{dcolumn}
\usepackage{bm}
\usepackage{graphics}
\usepackage{epsfig,color,soul}

\newcommand{\be}{\begin{equation}}
\newcommand{\ee}{\end{equation}}
\newcommand{\bea}{\begin{eqnarray}}
\newcommand{\eea}{\end{eqnarray}}

\begin{document}
\title{Theoretical investigation of the behavior of $CuSe_2O_5$ compound in high magnetic fields}
\author{Z. Saghafi, J. Jahangiri, S. Mahdavifar, H. Hadipour, S. Farjami Shayesteh}
\affiliation{ $^{1}$Department of Physics, University of
Guilan, 41335-1914, Rasht, Iran}
\date{\today}
\begin{abstract}
Based on analytical and numerical approaches, we investigate thermodynamic properties of $CuSe_2O_5$ at high magnetic fields which is a candidate for the strong intra-chain interaction in quasi one-dimensional (1D) quantum magnets.  Magnetic behavior of the system can be described by the 1D spin-1/2 XXZ model in the presence of the Dzyaloshinskii-Moriya (DM) interaction.  Under these  circumstances, there is one quantum critical field in this compound. Below the quantum critical field the spin chain system is in the gapless Luttinger liquid (LL) regime, whereas above it one observes a crossover to the gapped saturation magnetic phase.
  Indications on the thermodynamic curves confirm the occurrence of such a phase transition.  The main characteristics of the LL phase are gapless and spin-spin  correlation functions decay algebraic. The effects of zero-temperature quantum phase transition are observed even at rather high temperatures in comparison with the counterpart compounds. In addition, we calculate the Wilson ratio in the model. The Wilson ratio at a fixed temperature remains almost independent of the field in the LL region. In the vicinity of the quantum critical field, the Wilson ratio increases and exhibits anomalous enhancement.
  \end{abstract}
\pacs{75.10.Jm; 75.10.Pq}
\maketitle

\section{Introduction}\label{sec1}

From the theoretical point of view, the behavior of quantum magnets usually is determined by the Heisenberg model of interaction between spins. Moreover, it is known that a lack of symmetry in some quantum magnets leads to consider an extra interaction which is known as the Dzyaloshinskii-Moriya (DM) interaction\cite{Dzyaloshinskii58, Moriya60}. In fact, the DM interaction arises from the spin-orbit coupling and generates many surprising characteristics.

Experimentally, some quasi-1D antiferromagnetic systems are known to be described by the DM interaction\cite{Dender96, Dender97, Kohgi01, Fulde95, Oshikawa99, Tsukada01, Grande75, Povarov11, Greven95, Yildirim95, Katsumata01}. Recently a compound with the chemical formula, $CuSe_2O_5$, is known as a candidate for the quasi-1D quantum magnets with the DM interaction\cite{Meunier76, Kohn80, Becker06, Janson09, Choi11, Herak11, Herak13}.  DM interaction in a realistic model with multiorbitals on the ligand oxygen ions is the origin of anisotropic behavior which exists in this compound.\cite{Koshibae}
First, the structure of $CuSe_2O_5$ is characterized by chains with $Cu$ ions aligned along the $c$ axis of a monoclinic lattice\cite{Meunier76}. The mentioned crystal structure is reinvestigated\cite{Becker06} in $2006$ and to much higher precision is confirmed. By investigating the magnetic susceptibility, a strong intra-chain antiferromagnetic exchange interaction is revealed\cite{Kohn80}. In addition to the mentioned intra-chain exchange coupling,  it is argued that a non-frustrated very weak inter-chain coupling exists between the  nearest-neighbor spins\cite{Janson09}. This interaction causes a 3D antiferromagnetic ordering in the range of temperature below $17~K$. Using the Raman scattering study, it is confirmed that $CuSe_2O_5$ is characterized by a moderate, non-frustrated inter-chain coupling\cite{Choi11}. Using the electron spin resonance method, the low-temperature behavior is also studied\cite{Herak11}. It is found that the symmetric anisotropic exchange and the antisymmetric DM interaction are almost the same in this system. In a very recent work, using the bulk magnetization, neutron diffraction, muon spin relaxation and antiferromagnetic resonance measurements the magnetic behavior of this compound is studied. The long-range Neel order below $T_N=17~K$ is characterized\cite{Herak13}.

Theoretically, Using the high-temperature series expansion\cite{Janson00} and the Bethe-ansatz approach the temperature behavior of this compound is studied\cite{Janson09, Herak11, Herak13}. A very good agreement with the experimental results is found in the temperature region $T>T_N=17~K$ where the thermal fluctuations are dominant to the effect of the inter-chain exchange interaction. These theoretical studies are limited to a range of the magnetic field less than $5~T$. Since the intra-chain exchange interaction is very strong in this compound ($J=157~K$), the saturation field is very high and experimentally is not available now.

 Since the physics of the mentioned compound can be obtained by the 1D spin-1/2 antiferromagnetic Heisenberg model with added the DM interaction, the investigation of the high-field behavior is possible from the theoretical point of view. Therefore, in this paper, by using the analytical fermionization approach  and the numerical stochastic series expansion QMC method (the ALPS\cite{Albu07} code), we draw a theoretical clear picture of the probable behavior of the compound  $CuSe_2O_5$ in very high magnetic fields. Specially, we study the Wilson ratio in this compound. Since, recently the Wilson ratio on a gapped spin-1/2 two-leg ladder has been measured in an experiment\cite{Ninio12}, we suggest that the $CuSe_2O_5$ compound is a very good candidate for measuring the Wilson ratio. In addition, this is an opportunity for expanding the knowledge on the universal nature  of 1D quantum liquids.

The paper is organized as follows. In the forthcoming section we introduce the model and map it into an effective 1D spin-1/2 XXZ model in the presence of a magnetic field. Using the fermionization approach we diagonalize the Hamiltonian.  In section III, we present our analytical and numerical results and compare them. Finally, we conclude and summarize our results in section IV.

\section{THE MODEL}\label{sec2}

The Hamiltonian of the 1D spin-1/2 antiferromagnetic Heisenberg model with added DM interaction in a magnetic field is written as
\begin{eqnarray}
{\cal H}&=&J\sum_{j=1}^{N}(\overrightarrow{S}_{j}\cdot\overrightarrow{S}_{j+1})+ \sum_{j=1}^{N} (-1)^{j} \overrightarrow{D}.( \overrightarrow{S}_{j}\times \overrightarrow{S}_{j+1})  \nonumber\\
&-&g \mu_B H\sum_{j=1}^{N}S_{j}^{z},
\label{Hamiltonian s}
\end{eqnarray}
 where $S_{j}$ is the spin-1/2 operator on the $j$-th site, $J>0$ denotes the exchange coupling constant, $H$ is the applied magnetic field and $\overrightarrow{D}=D \hat{z}$ is known as the DM vector. The exchange coupling and the DM vector are measured $J=157~K$, $D=0.05~J$ respectively in the $CuSe_2O_5$ compound. At first, by considering a rotation\cite{Oshikawa97} around the $Z$ axis as $S_{j}^{\pm} \longrightarrow \widetilde{S}_{j}^{\pm} \exp(\mp i\alpha)$ in which $\tan{\alpha}=\frac{-D}{J}$, the Hamiltonian is transformed to the 1D anisotropic spin-1/2 XXZ model in a longitudinal magnetic field

\begin{eqnarray}
{\cal
 H}&=&\tilde{J}\sum_{j=1}^{N}(\widetilde{S}_{j}^{x}\widetilde{S}_{j+1}^{x}+\widetilde{S}_{j}^{y}\widetilde{S}_{j+1}^{y}+\frac{J}{\tilde{J}}\widetilde{S}_{j}^{z}\widetilde{S}_{j+1}^{z})\nonumber\\
  &-&g \mu_B H\sum_{j=1}^{N}\widetilde{S}_{j}^{z},
\label{Hamiltonian-tf}
\end{eqnarray}
 with an effective exchange interaction $\tilde{J}=\sqrt{J^{2}+D^{2}}$. Since the exchange anisotropy is less than the $XY$ anisotropy ($J<\tilde{J}$), the ground state of the system is in the Luttinger liquid (LL) phase at zero temperature\cite{Takahashi99}. By increasing the magnetic field from zero, up to the critical field  $H_c= \frac{\tilde{J}+J}{g \mu_B}$, the ground state remains in the LL phase where a  quantum phase transition into the ferromagnetic phase with saturation magnetization along the field will occur.

Theoretically, the energy spectrum is needed to study the thermodynamic behavior of the model. An efficient method for diagonalizing is the analytical fermionization technique. At the second step, by implementing the Jordan-Wigner transformations,

\begin{eqnarray}
\widetilde{S}_{j}^{z}&=&a_{j}^{\dagger}a_{j}-\frac{1}{2}, \\ \nonumber
\widetilde{S}_{j}^{+}&=&a_{j}^{\dagger} \exp(i\pi\sum_{l<j}a_{l}^{\dagger}a_{l}),\\ \nonumber
\widetilde{S}_{j}^{-}&=&a_{j} \exp(-i\pi\sum_{l<j}a_{l}^{\dagger}a_{l}),
\end{eqnarray}
the transformed Hamiltonian is mapped into a 1D model of interacting spinless fermions,
\begin{eqnarray}
{\cal H}_{f}&=& \frac{N g \mu_B H}{2}+\frac{N J}{4}+
\frac{\widetilde{J}}{2} \sum_{j} (a^{\dag}_{j}a_{j+1}+a^{\dag}_{j+1}a_{j})\nonumber \\
&+& J \sum_{j} a^{\dag}_{j}a_{j}  a^{\dag}_{j+1}a_{j+1}\nonumber \\
&-&(J+g \mu_B H) \sum_{j} a^{\dag}_{j}a_{j}~.
\end{eqnarray}
Then, using the Wick's theorem, the fermion interaction term is decomposed by some order parameters which are
related to the spin-spin correlation functions as\cite{Dmitriev02}
\begin{eqnarray}
\gamma_1&=&\langle a^{\dag}_{j}a_{j} \rangle , \nonumber \\
\gamma_2&=&\langle a^{\dag}_{j}a_{j+1} \rangle ,\nonumber \\
\gamma_3&=&\langle a^{\dag}_{j}a^{\dag}_{j+1}\rangle .
\end{eqnarray}
By utilizing these order parameters and performing a Fourier transformation as $a_{j} = \frac{1}{\sqrt{N}} \sum ^{N} _{j=1} e^{-ikj} a_{k}$, and also using the following unitary transformation
\begin{eqnarray}
a_{k}=cos(k) \beta_k-i sin(k) \beta^{\dag}_{k},
\end{eqnarray}
the diagonalized Hamiltonian is given by
\begin{eqnarray}
{\cal H}_{f}=\sum_{k=-\pi}^{\pi}\varepsilon(k) (\beta_{k}^{\dagger} \beta_{k}-\frac{1}{2}).
\label{Hamiltonian d}
\end{eqnarray}
Where the dispersion relation is
\begin{eqnarray}
\varepsilon(k) &=&  \sqrt{a(k)^2+4b(k)^2}, \nonumber\\
a(k)&=&(\widetilde{J}-2 \gamma_2 J) cos(k)+2 \gamma_1 J-(g \mu_B H+J), \nonumber\\
b(k)&=& J \gamma_3 sin(k).
\end{eqnarray}
In order to solve ${\cal H}_{f}$, the following
 equations should be satisfied self-consistently
\begin{eqnarray}
\gamma_1&=&1+\frac{1}{2 \pi} \int_{-\pi}^{\pi} dk~\frac{a(k)}{\varepsilon(k)}(\frac{1}{1+e^{\varepsilon(k)/K_B T}}-\frac{1}{2}),  \nonumber  \\
\gamma_2&=&\frac{1}{2 \pi} \int_{-\pi}^{\pi} dk~ \frac{a(k)}{\varepsilon(k)}(\frac{1}{1+e^{\varepsilon(k)/K_B T}}-\frac{1}{2}) cos(k), \nonumber \\
\gamma_3&=&\frac{1}{2 \pi} \int_{-\pi}^{\pi} dk~ \frac{b(k)}{\varepsilon(k)}(\frac{1}{1+e^{\varepsilon(k)/K_B T}}-\frac{1}{2}) sin(k).
\label{self}
\end{eqnarray}
For this purpose, we have written a computational code to solve the self consistent equations. Using our code, first we find a set of $\gamma_i, i=1, 2, 3$ with fixing the values of the exchange coupling, the DM vector and the magnetic field. Then, by including $\gamma_i, i=1, 2, 3$ into the dispersion relation, the energy spectrum of the model is obtained.
Finally the thermodynamic functions such as the free energy, the specific heat, the magnetization and the susceptibility are obtained as
\begin{eqnarray}
f&=&-\frac{K_B T}{2 \pi} \int_{-\pi}^{\pi} dk~ln(2 cosh(\frac{\varepsilon(k)}{2 K_B T})), \nonumber \\
C_V&=&-\frac{1}{8 \pi K_B T^2} \int_{-\pi}^{\pi} dk~(\varepsilon(k)^{2} sech^{2}(\frac{\varepsilon(k)}{2 K_B T})), \nonumber \\
m&=&\frac{1}{2 \pi} \int_{-\pi}^{\pi} dk~(\frac{a(k)}{\varepsilon(k)}(\frac{1}{1+e^{\varepsilon(k)/K_B T}})-\frac{a(k)}{2 \varepsilon(k)}), \nonumber \\
\chi&=&\frac{\partial m}{\partial h}.
\end{eqnarray}

 The analytical fermionization approach is used for investigating the behavior of the system from a theoretical point of view.

\section{Results}\label{sec4}

\begin{figure}[t]
\centerline{\psfig{file=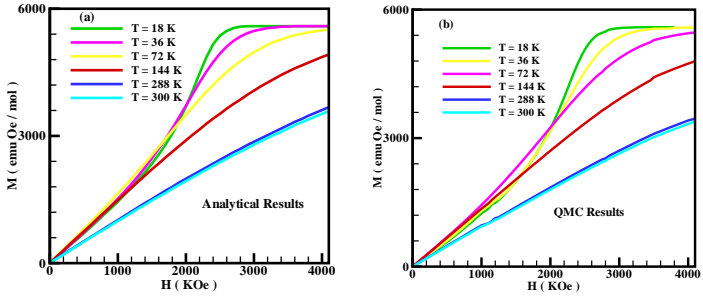,width=3.65in}}
\caption{Magnetization versus magnetic field at zero temperature showing a critical field at $H_c=2300~KOe$ (a) fermionization approach  (b) SSE QMC simulation. }\label{fig1}
\end{figure}

\begin{figure}[t]
\centerline{\psfig{file=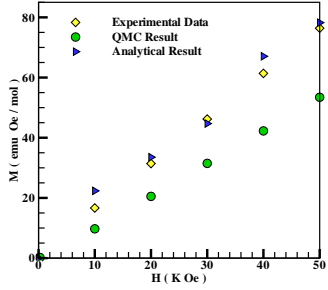,width=1.8in}}
 \caption{Magnetization versus magnetic field at temperature T = 4.2 K which is coincident to experimental data. Fermionization approach and SSE QMC simulation has been carried out. }\label{fig2}
\end{figure}

In this section we present our results regarding the thermodynamic behavior of the 1D spin-1/2 XXZ model in the presence of a longitudinal magnetic field in with DM interaction. As mentioned before, the field of our study is the $CuSe_2O_5$ compound at high magnetic fields. To this end, we use the experimental data in our calculation, for compatibility between theory and experiment. The maximum size of spin chains in QMC simulations is $100$ particles. The QMC simulation is performed for spin chains under the periodic boundary condition with the maximum
3$\times$10$^{6}$ equilibration sweeps and 6$\times$10$^{6}$ measurement steps. It is believed that the simulation approaches are very good to find the results in cases where the performing of a real experiment is impossible. Since, there are not any real experimental results for the behavior of the $CuSe_2O_5$ compound in high magnetic field, we have used the QMC simulation. We have also solved the integral equations (Eq.~(\ref{self})) numerically to find the energy spectrum of the model in the fermionization approach.
Figs.~\ref{fig1}(a) and 1(b) show the magnetization versus the longitudinal magnetic field at different values of the temperature by using fermionization approach and SSE QMC simulations respectively. At very low temperature, both of them show that the magnetization starts from zero, and it confirms that the excitation spectrum of spin chain is in the gapless LL region. By increasing the magnetic field, the magnetization increases very slowly because of the strong exchange interaction. Thus, the very large value of the field  $H_c=2300~KOe$ has the role of quantum critical field and in the region of $H>H_c$ the magnetization will be saturated. The spin chain system undergoes a crossover from the gapless LL phase to the gapped saturation magnetic phase. As is seen, a very good agreement exists between the QMC simulation and fermionization approach results. By increasing the temperature, the saturation plateau will be destroyed due to the thermal fluctuations.  Experimentally the magnetization process was studied in the temperature $T=4.2~K$ and the magnetic fields up to $50 KOe$\cite{Herak13}.  Fig.~\ref{fig2} shows our analytical and numerical results together with the experimental data\cite{Herak13}. It is clearly seen that, at $T=4.2~K$ our theoretical results have a very good agreement with the experimental data.

\begin{figure}[t]
\centerline{\psfig{file=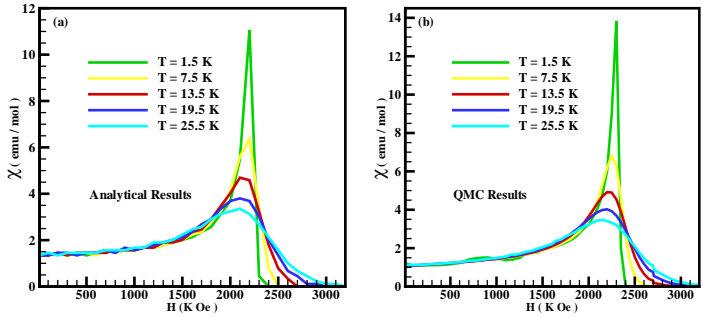,width=3.65in}}
 \caption{Molar susceptibility versus magnetic field at different temperatures. The asymptotical behavior in the vicinity of zero temperature is an indication of cross-over to a saturation magnetic phase (a) fermionization approach (b) SSE QMC simulation.}\label{fig3}
\end{figure}

\begin{figure}[t]
\centerline{\psfig{file=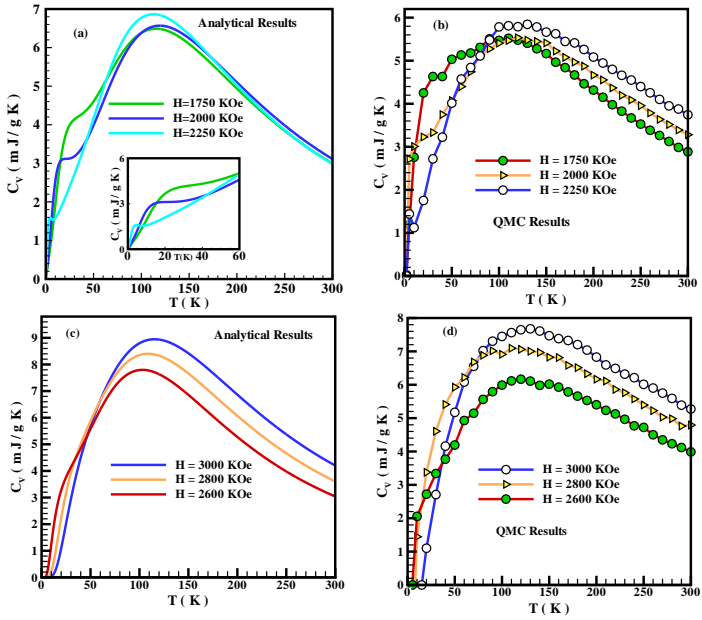,width=3.65in}}
\caption{Magnetic specific heat versus temperature below critical magnetic field $H_c=2300~KOe$ (a) fermionization approach (b) SSE QMC simulation. (c) Magnetic specific heat versus temperature above critical magnetic field $H_c=2300~KOe$ with fermionization approach (d) SSE QMC simulation.}\label{fig4}
\end{figure}

\begin{figure}[t]
\centerline{\psfig{file=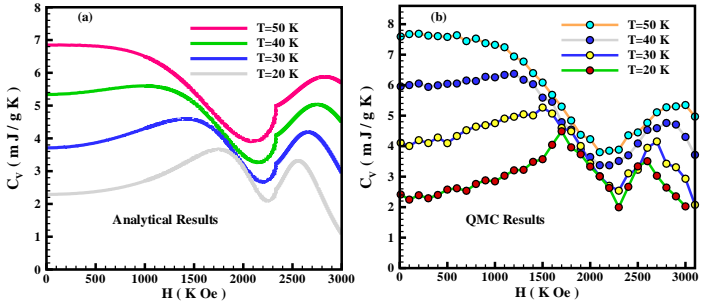,width=3.65in}}
\caption{Magnetic specific heat versus magnetic field at different temperatures (a) fermionization approach (b) SSE QMC simulation. The quantum critical field shows its effect at rather high temperatures.}\label{fig5}
 \end{figure}

 Figs.~\ref{fig3} (a) and 3(b) illustrate the molar susceptibility versus the magnetic field with both SSE QMC and fermionization approaches. At quite low temperature ($T = 1.5 K$), by increasing the magnetic field up to $2300~KOe$ a sharp peak emerges in $\chi(H)$, which is an indication of the quantum critical field. Due to the very large exchange interaction, as depicted in Figs.~\ref{fig3} (a) and 3(b),  at magnetic fields nearly up to $2000~KOe$ the molar susceptibility is the same within all the considered range of temperature which is one of the important signatures of the LL phase. By more increasing of the magnetic field up to the quantum critical field in the region of LL phase the molar susceptibility decreases with increasing temperature. As expected, enhancing of the temperature causes the sharp peak to fade. In other words, the effects of quantum fluctuations disappear and thermal fluctuations are dominant. The behavior of molar susceptibility in the gapped saturation magnetic phase is different from the gapless LL phase. In other words, we observe that the molar susceptibility increases with enhancing of temperature.
To complete our study, we calculate the specific heat for different values of the temperature and the magnetic field. As usual, figures have been plotted with two numerical and analytical results. Figs.~\ref{fig4}(a)-4(d) show the magnetic specific heat versus the temperature. Below the critical field $H<H_c$, with lowering the temperature, there is a broad peak about $100~K$ which is called a Schottky peak, then with more decreasing temperature we observe a second peak in the specific heat and comes down to zero linearly, indicating the existence of the LL gapless regime. By increasing the magnetic field and near to the quantum critical field the height of the second peak decreases and then vanishes. Figs.~\ref{fig4} (c) and 4(d) have been plotted for $H>H_c$. Both QMC simulation and fermionization approaches give us the same results. Above the  quantum critical field $H>H_c=2300~KOe$, a crossover occurs into the gapped saturation magnetic phase and there is no trace of the second peak. On the other hand, $C_V(T)$ remains zero up to a threshold temperature, which is known as an indication of the energy gap in the saturation magnetic field.
 Next, we consider the specific heat versus the magnetic field $C_V(H)$ at a fixed temperature.
 As illustrated in Figs.~\ref{fig5} (a) and 5(b) the curve of $C_V (H)$ has been plotted at various fixed temperatures $T=20~K, 30~K , 40~K$ and $50~K$. It is clearly seen from this figure that the specific heat at low temperature, shows two-peaks structure on both sides of the quantum critical point in good agreement with results obtained for the Kagome antiferromagnet chains\cite{Zhitomirsky04} and Ising model\cite{Hasanzadeh14}. Within the temperature mentioned above, there is a minimum at $H_c=2300~KOe$, which is an indication of the existence of the zero-temperature quantum phase transition. This minimum fades with enhancing temperature, because of thermal fluctuations. The signature of existence of zero-temperature quantum phase transition in the curve of $C_V(H)$ even in this range of temperature (up to $100~K$) is because of the strong exchange interaction in this compound. Based on these results, it seems that quantum correlations\cite{Amico08} as the entanglement in the $CuSe_2O_5$ compound can be detected experimentally at high temperatures which can be considered as an open problem.
In addition, we have focused on the theoretical evaluation of the Wilson ratio in the $CuSe_2O_5$ compound. It is known that the Wilson ratio\cite{Wilson75},
\begin{eqnarray}
R_{W}=\frac{4}{3} (\frac{\pi k_B}{g \mu_B})^{2} \frac{\chi}{C_{v}/T},
\label{Hamiltonian d}
\end{eqnarray}
is a crucial parameter for characterizing the LL region. In this relation, $\chi$ is the magnetic susceptibility and $C_V$ is the specific heat. In principle, the Wilson ratio quantifies the interaction effect and the spin fluctuations which enhance the magnetic susceptibility. It was illustrated that $R_W$ is $2$ for the spin-1/2 Kondo lattice in the single-impurity limit\cite{Wilson75}. It will be $1$ for a system of noninteracting electrons in a metal\cite{Hewson97} and independent of the temperature. Since the spin fluctuations are enhanced in strongly correlated systems, $R_W>1$ is expected . Since the Wilson ratio has recently been measured in experiments on a gapped spin-1/2 Heisenberg ladder\cite{Ninio12}, the study of the Wilson ratio in the 1D spin systems has also attracted much interest. At zero temperature,  $R_W$ exactly equals 2 for an isotropic antiferromagnetic spin-1/2 chains\cite{Johnston00}. In this system, $R_W$ also depends on the temperature. Recently the effect of the polarization on the Wilson ratio of Fermi gases in 1D has been studied\cite{Guan13-1, Guan13-2, Guan14}. It is found that $R_W$ exhibits anomalous enhancement at critical points.

\begin{figure}[t]
\centerline{\psfig{file=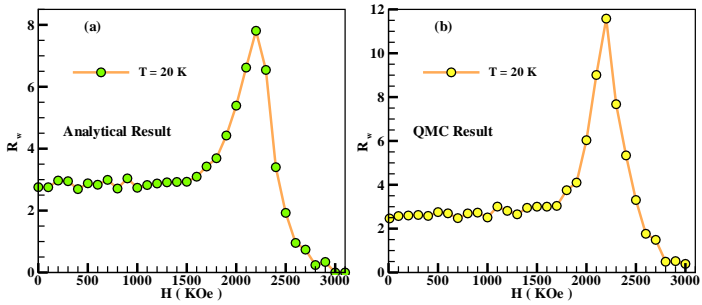,width=3.65in}}
\caption{The Wilson ratio as a function of the magnetic field at fixed temperature $T=20 K$(a) fermionization approach (b) SSE QMC simulation.}\label{fig6}
 \end{figure}

\begin{figure}[t]
\centerline{\psfig{file=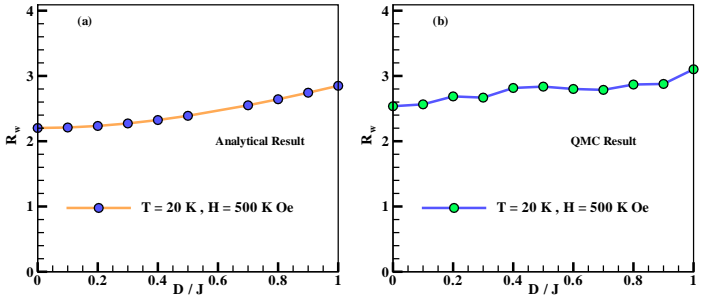,width=3.65in}}
\caption{The Wilson ratio as a function of the DM vector at fixed temperature $T=20 K$ and magnetic field $H=500 KOe$ (a) fermionization approach (b) SSE QMC simulation.}\label{fig7}
 \end{figure}

Here we have calculated the Wilson ratio in our model. In Figs.~\ref{fig6} (a) and 6(b), $R_W$ has been plotted as a function of the magnetic field at a fixed temperature $T=20 K$. The special value of $20 K$ has been selected since $CuSe_2O_5$ compound behaves as a quasi-1D system at temperatures higher than $T_N=17 K$. Our analytical and numerical results show that the Wilson ratio is $2<R_W<3$ and  independent of the magnetic field in the region of $H<H_c$, where the LL phase is expected. An anomalous enhancement of the Wilson ratio in the vicinity of the quantum critical point is seen. This phenomenon has also been observed near the metal-insulator transition in simulations of a three dimensional quantum spin liquid\cite{Chen13}. Here, this anomalous divergence is mainly an indication of the quantum critical point. In addition, this behavior confirms that the susceptibility of the system increases in the vicinity of the quantum critical point much faster than the specific heat. In other words, the critical exponent of the susceptibility should be larger than the critical exponent of the specific heat.
It is interesting, what happens when we change the value of DM interaction, however the value of measured DM vector is constant ($D = 0.05 J$). Since the transformed Hamiltonian (Eq. (2)) is a 1D spin-1/2 XXZ model, we expect that by increasing the DM interaction the
model remains in the LL phase and no quantum phase transition occurs at zero temperature. Among the thermodynamic functions, the evaluation of the behavior of the Wilson ratio in respect to the DM vector is interesting. In Figs.~\ref{fig7} , the Wilson ratio has been plotted as a function of the DM interaction at a fixed temperature, T = 20K and H = 500KOe. As is seen, $R_W$ remains almost constant and independent of the DM vector in this region.
We expect this behavior, since the system remains in the LL phase by increasing the DM vector.

Finally, using the mean field approximation\cite{Schulz96} we have estimated  the critical Neel temperature in the $CuSe_2O_5$ compound. We have approximated the $CuSe_2O_5$ compound with  a system of coupled parallel XXZ chains where the exchange interaction among neighboring chains is $J_{in}=0.1 J=15.7 K$. The condition for determining $T_N$ is $\frac{z J_{in}}{(g\mu_B)^2} \chi(T)=1$, where $z$ is the coordination number of the lattice. Using this relation, we found the Neel temperature $T_N\simeq22~K$, which is in good agreement with the experiment.

\section{Conclusion}\label{sec2}

In summary,  we have investigated the high magnetic field behavior of the quasi-1D  $CuSe_2O_5$ compound. It has been suggested that the behavior of this compound can be obtained by the 1D spin-1/2 Heisenberg model with the Dzyaloshinskii-Moriya (DM) interaction. At the first step, we have mapped the Hamiltonian of the system into an effective 1D spin-1/2 XXZ model. Then, analytically the effective Hamiltonian is diagonalized  using the fermionization approach. Using the energy spectrum, the thermodynamic functions are calculated. In addition, the numerical SSE QMC simulation is also applied and the same thermodynamic functions are calculated for finite chain systems up to $N=100$ spins.

Results show that the system is in the LL phase in the presence of the external magnetic fields up to $2300 KOe$. With further increase of the magnetic field, the system undergoes a phase transition into the saturation magnetic phase. The molar susceptibility is independent of the temperature in the LL phase. Also, specific heat versus temperature shows two-peaks in the LL region, which is a very important signature of the LL phase. In addition, in the figure of the specific heat versus magnetic field, two-peaks structure on both sides of the quantum critical point can be observed in high values of temperature $\simeq 50K$. The mentioned  behavior suggests  that the quantum correlations in this compound should be very strong and their effects can be observed at high temperature region. In the topic of "quantum entanglement"\cite{Amico08} and its applications, having the quantum correlations at high temperature region is a challenge. Our results show that the $CuSe_2O_5$ compound can be a very good candidate for studying the entanglement in the experiment.

We also tried to find an estimation of the Wilson ratio's order in the $CuSe_2O_5$ compound. Our results show that the Wilson ratio should be in the range $2<R_W<3$. On the other hand, it remains almost constant up to very high magnetic fields $\simeq 2000 KOe$.

\section{acknowledgment}
It is our pleasure to thank M. Herak for very useful comments.
\vspace{0.3cm}


\end{document}